\documentstyle[12pt]{article}
\begin{document}
\def\beq{\begin{equation}}
\def\eeq{\end{equation}}
\def\bea{\begin{eqnarray}}
\def\eea{\end{eqnarray}}
\def\ve{\vert}
\def\vel{\left|}
\def\ver{\right|}
\def\nnb{\nonumber}
\def\ga{\left(}
\def\dr{\right)}
\def\aga{\left\{}
\def\adr{\right\}}
\def\rar{\rightarrow}
\def\nnb{\nonumber}
\def\la{\langle}
\def\ra{\rangle}
\def\lla{\left<}
\def\rra{\right>}
\def\ba{\begin{array}}
\def\ea{\end{array}}
\def\tep{$B \rar K \ell^+ \ell^-$}
\def\tepm{$B \rar K \mu^+ \mu^-$}
\def\tept{$B \rar K \tau^+ \tau^-$}
\def\ds{\displaystyle}



\def\bos{\lower 0.5cm\hbox{{\vrule width 0pt height 1.2cm}}}
\def\boss{\lower 0.35cm\hbox{{\vrule width 0pt height 1.cm}}}
\def\aaa{\lower 0.cm\hbox{{\vrule width 0pt height .7cm}}}
\def\dol{\lower 0.4cm\hbox{{\vrule width 0pt height .5cm}}}


\title{ {\Large {\bf $g_{K\Lambda N}$ and $g_{K\Sigma N}$ coupling
constants in light cone QCD sum rules} } }

\author{\vspace{1cm}\\
{\small T. M. Aliev \thanks
{e-mail: taliev@rorqual.cc.metu.edu.tr}\,\,,
M. Savc{\i} \thanks
{e-mail: savci@rorqual.cc.metu.edu.tr}} \\
{\small Physics Department, Middle East Technical University} \\
{\small 06531 Ankara, Turkey} }
\date{}

\begin{titlepage}
\maketitle
\thispagestyle{empty}

\begin{abstract}
\baselineskip  0.7cm
The strong coupling constants $g_{K\Lambda N}$ and  $g_{K\Sigma N}$ for 
the structure 
$\sigma_{\mu\nu} \gamma_5$ are calculated within light cone QCD sum rules. A
comparison of our results on these couplings with predictions from
traditional QCD sum rules is presented.
\end{abstract}

\vspace{1cm}
\end{titlepage}

\section{Introduction}
In understanding the dynamics of the kaon nucleon scattering or photo--kaon
production in nucleon, it is important to know the hadronic coupling
constants involving the kaons. Among them, $g_{K\Lambda N}$ and 
$g_{K\Sigma N}$ are the most relevant coupling constants.
Phenomenological models for determination of these constants from
kaon--nucleon scattering and from the kaon photo--production, involve many
unknown parameters (see for example \cite{R1} and references therein).   
Therefore any prediction about these constants is strongly model dependent
and suffers from large uncertainties. For this reason a quantitative
calculation of the $g_{KYN}~(Y=\Lambda$ or $\Sigma)$ coupling constants 
with a tractable and reliable theoretical approach is needed. 

It is widely accepted that QCD is the underlying theory of the strong
interactions. In the typical hadronic scale the strong coupling constant 
$\alpha_s(\mu=m_{had})$ becomes large and QCD is nonperturbative. For this
reason calculation of $g_{KYN}$ is related to the nonperturbative sector of
QCD, and some kind of nonperturbative approach is needed 
for determination of the above--mentioned coupling constants. Among various 
nonperturbative methods, QCD sum rules \cite{R2} is a powerful one. This 
method is based on the short distance OPE of vacuum vacuum correlation 
function in terms of condensates.  For the processes involving light mesons
$\pi,~K$ or $\rho$, there is an alternative method to the traditional QCD
sum rules, namely, light cone QCD sum rules \cite{R3}. In this approach the
expansion of the vacuum--meson correlator is performed near the light cone
in terms of the meson wave functions. The meson wave functions are defined
by the matrix elements of non--local composite operators sandwiched between
the meson and vacuum states and classified by their twists, rather
than dimensions of the operators, as is the case in the traditional sum
rules (more about application of light--cone QCD sum rules can be found in
\cite{R5}--\cite{R12} and references therein).
  
In this work we employ light cone QCD sum rules method to extract the 
coupling constants $g_{KYN}$. These coupling constants were investigated in framework 
of the traditional QCD sum rules method in \cite{R1,R13} for the structure 
$\not\!q \gamma_5$, and for the structure $\sigma_{\mu\nu} \gamma_5$ in
\cite{R14}. The discrepancy
between the results of these works makes it necessary to perform independent
calculations in determining the coupling constants $g_{KYN}$. 
In the present article we restrict ourselves to the consideration of the
structure $\sigma_{\mu\nu} \gamma_5$ whose choice is dictated by the
following reason. In \cite{R15,R16} it was pointed out that there is
coupling scheme dependence for the structures $\gamma_5,\not\!q \gamma_5$,
(i.e., dependence on the pseudoscalar or pseudovector forms of the effective 
interaction Lagrangian of pion with hadrons in the phenomenological part,
have been used), while the structure $\sigma_{\mu\nu} \gamma_5$ is shown to
be independent of any coupling schemes.

In order to calculate the coupling constants $g_{KYN}$ we start with the
following two--point function
\bea
\Pi(p,p_1,q) = \int d^4x \, e^{ipx} \lla 0 \vel \mbox{\rm T} \eta_Y (x)
\bar \eta_N (0) \ver K(q) \rra~, 
\eea
where $p$ and $\eta_Y$ are the four--momentum of the hyperon and its
interpolating current, respectively, $\eta_N$ is the nucleon interpolating
current, $q$ is the four--momentum of $K$ meson. The interpolating currents
for $\Lambda$, $\Sigma$ and $N$ are \cite{R17,R18}
\bea
\eta_\Lambda &=& \sqrt{\frac{2}{3}}\, \epsilon_{abc} \left[
\ga u_a^T C \gamma_\mu s_b \dr \gamma_5 \gamma^\mu d_c -
\ga d_a^T C \gamma_\mu s_b \dr \gamma_5 \gamma^\mu u_c \right]~, \nnb \\
\eta_{\Sigma^0} &=& \sqrt{2} \, \epsilon_{abc} \left[
\ga u_a^T C \gamma_\mu s_b \dr \gamma_5 \gamma^\mu d_c +
\ga d_a^T C \gamma_\mu s_b \dr \gamma_5 \gamma^\mu u_c \right]~, \nnb \\
\eta_N &=& \epsilon_{abc} \ga u_a^T C \gamma_\mu u_b \dr 
\gamma_5 \gamma^\mu d_c~,
\eea
where $s,~u$ and $d$ are strange, up and down quark fields, respectively and
$C$ is the charge conjugation operator, $a,~b$ and $c$ are the color indices. 

As has already been mentioned, it was pointed out in \cite{R14,R15} that a
better determination of $g_{\pi NN}$ can be done by the structure 
$\sigma_{\mu\nu}\gamma_5$, since this structure is independent of the
effective models employed in the phenomenological part. This fact motivates us
to calculate  $g_{NYK}$ in this structure.

Using the Lorentz, parity and charge conjugation invariance, 
$T(p,p_1,q)$ can be represented in the following general form
\bea
T(p,p_1,q) &=& T_1(p^2,p_1^2,q^2) \gamma_5 +
T_2(p^2,p_1^2,q^2) \not\!q \gamma_5 \nnb \\
&&+ T_3(p^2,p_1^2,q^2) \not\!P \gamma_5 +
T_4(p^2,p_1^2,q^2) \sigma_{\mu\nu} \gamma_5 p^\mu q^\nu~,
\eea
where $q=p-p_1$, $P=(p+p_1)/2$~. 

On the phenomenological part,
these different Dirac structures are obtained by saturating correlator (1)
of the $Y$ and $N$ states
\bea
T=
\frac{\lla Y(p) \ve K(q) N(p_1) \rra \, \lla 0 \ve \eta_Y(x) \ve Y(p) \rra
\, \lla N(p_1)\ve \bar \eta_N (0) \ve 0 \rra}
{(p^2-m_Y^2) (p_1^2 - m_N^2)}~.  
\eea
The matrix elements in Eq. (4) are defined in the following way
\bea
\lla 0 \vel \eta_Y(x) \ver Y(p) \rra &=& \lambda_Y u(p)~, \nnb \\
\lla N(p_1) \vel \bar \eta_N(x) \ver 0 \rra &=& \lambda_N \bar u(p_1)~,\nnb \\
\lla Y(p)\ve K(q) N(p_1) \rra &=& g_{KYN} \, \bar u(p) \gamma_5 u(p_1)~.
\eea
Substituting Eq. (5) in Eq. (4), we get
\bea
T=\frac{g_{KYN} \lambda_Y \lambda_N}{(p^2-m_Y^2) (p_1^2 - m_N^2)} \,
\ga \not\!p +m_Y \dr \gamma_5 \ga \not\!p_1 + m_N \dr +
\mbox{\rm higher resonances}~, \nnb
\eea
which can be written as
\bea
T&=&\frac{g_{KYN} \lambda_Y \lambda_N}{(p^2-m_Y^2) (p_1^2 - m_N^2)} \,
\Big[ \ga m_Y m_N - p p_1 \dr \gamma_5 +
\frac{m_Y + m_N}{2} \, \not\!q \gamma_5 \nnb \\
&& -  \frac{m_Y -m_N}{2} \, \not\!P \gamma_5  -
i \sigma_{\alpha\beta} p_\alpha q_\beta \gamma_5 \Big]~.
\eea
Choosing the structure $\sigma_{\alpha\beta} \gamma_5$ as the physical part,
we have 
\bea
T^{phys} = -i\, \frac{g_{KYN} \lambda_Y \lambda_N}
{(p^2 - m_Y^2 ) ( p_1^2 - m_N^2)} \, p_\alpha q_\beta~.
\eea
Let us now turn our attention to the theoretical part of the correlator (1).
From Eq. (1) we immediately get
\bea
T&=& \alpha \int dx\,e^{ipx} \Big\{ - 4 \gamma_5 \gamma_\mu i {\cal S}
\gamma_\nu \gamma_5 \lla 0 \vel \bar u(0) \gamma_\nu 
{\cal C} i {\cal S}^T {\cal C}^{-1} \gamma_\mu 
s(x) \ver K(q) \rra \nnb \\
&& \mp \gamma_5 \gamma_\mu i {\cal S} \gamma_\nu \gamma_5 
\gamma_\mu i {\cal S}
\gamma_\nu  \gamma_5 \lla 0 \vel \bar u(0) \gamma_5 s(x) \ver K(q) \rra \nnb \\
&& \mp \gamma_5 \gamma_\mu i {\cal S} \gamma_\nu \gamma_5 \gamma_\rho
\gamma_\mu i {\cal S} \gamma_\nu \gamma_5
\lla 0 \vel \bar u(0) \gamma_\rho \gamma_5  s(x) \ver K(q) \rra \Big\}~,
\eea
where upper (lower)  sign corresponds to $\Lambda~(\Sigma)$ case, 
$\alpha=\sqrt{2/3}~(\sqrt{2})$ for $\Lambda~(\Sigma)$. In Eq. (8)  
${\cal S}$ is the full light quark propagator containing both 
perturbative and nonperturbative contributions,
\bea
i {\cal S} &=& i \, \frac{\not\!x}{2 \pi^2 x^4} - 
\left( \frac{\la \bar q q \ra}{12} + 
\frac{x^2 m_0^2}{192}\la \bar q q \ra \right) \nnb \\
&&- \, i \, \frac{g_s}{16 \pi^2} \int_0^1 du 
\left\{ \frac{\not\!x}{x^2} \sigma_{\alpha\beta} G^{\alpha\beta} (ux) -
4 i \frac{x_\alpha}{x^2} G^{\alpha\beta} \gamma_\beta \right\} 
+ \cdots
\eea
It follows from Eq. (8) that, in order to calculate the correlator function 
in QCD, the matrix elements of the nonlocal operators
between the vacuum and kaon states are needed. These matrix elements are
defined in terms of kaon wave functions, and up to twist four these wave
functions can be written as \cite{R6,R7}:
\bea
\lla 0 \vel \bar u (0) \gamma_\mu \gamma_5 s(x) \ver K(q) \rra &=& 
i f_\pi q_\mu \int_0^1 du \, e^{-iuqx} \big[ \varphi_K(u) + x^2 g_1(u) \big] 
\nnb \\
&&+f_\pi \ga x_\mu - \frac{x^2 q_\mu}{qx} \dr 
\int_0^1 du\, e^{-iuqx} g_2(u)~, \nnb \\ \nnb \\
\lla 0 \vel \bar u (0) i \gamma_5 s(x) \ver K(q) \rra &=&
f_K \mu_K \int_0^1 du \,e^{-iuqx} \varphi_p(u) ~, \nnb \\ \nnb \\
\lla 0 \vel \bar u (0) \sigma_{\alpha\beta} \gamma_5 s(x) \ver K(q) \rra &=&
\ga q_\alpha x_\beta - q_\beta x_\alpha \dr \frac{i f_K \mu_K}{6}
 \int_0^1 du \,e^{-iuqx} \varphi_\sigma(u)~, \nnb
\eea
\bea
\lefteqn{
\lla 0 \vel \bar u (0) \gamma_\mu \gamma_5 G_{\alpha\beta}(ux) s(x) \ver
K(q) \rra =}\nnb \\
&& \left[ q_\beta \ga g_{\alpha\mu} - \frac{x_\alpha q_\mu}{qx} \dr -
q_\alpha \ga g_{\beta\mu} - \frac{x_\beta q_\mu}{qx} \dr \right]
\int {\cal D}\alpha_i \,\varphi_\perp (\alpha_i) 
e^{-iqx(\alpha_1+ u \alpha_3)} \nnb \\
&&+ \frac{q_\mu}{qx} \ga q_\alpha x_\beta - q_\beta x_\alpha \dr 
\int {\cal D}\alpha_i \,\varphi_\parallel(\alpha_i) 
e^{-iqx(\alpha_1+ u \alpha_3)} \nnb \\ \nnb \\
\lefteqn{
\lla 0 \vel \bar u (0) \gamma_\mu i g \widetilde G_{\alpha\beta} (ux)
s(x) \ver K(q) \rra =}\nnb \\
&& \left[ q_\beta \ga g_{\alpha\mu} - \frac{x_\alpha q_\mu}{qx} \dr -  
q_\alpha \ga g_{\beta\mu} - \frac{x_\beta q_\mu}{qx} \dr \right]        
\int {\cal D}\alpha_i \,\tilde \varphi_\perp (\alpha_i)
e^{-iqx(\alpha_1+ u \alpha_3)} \nnb \\
&&+\frac{q_\mu}{qx} \ga q_\alpha x_\beta - q_\beta x_\alpha \dr 
\int {\cal D}\alpha_i \,\tilde \varphi_\parallel(\alpha_i) 
e^{-iqx(\alpha_1+ u \alpha_3)}~,
\eea
where 
\bea
\mu_K = \frac{m_K^2}{m_u+m_s}~~~~\mbox{\rm and}~~~~
\int {\cal D} \alpha_i \equiv \int d \alpha_1 \, d \alpha_2 \, \alpha_3 \, 
\delta(1-\alpha_1-\alpha_2-\alpha_3)~. \nnb 
\eea
Due to the choice of the gauge $x^\mu A_\mu(x) = 0$, the path 
ordered gauge
factor ${\cal P} e^{ i \ga g_s \int du\, x^\mu A_\mu(ux) \dr}$ has been
omitted. The wave function $\varphi_K(u)$ is the leading twist $\tau=2$,
$g_1(u),~\varphi_\perp,~\varphi_\parallel,~\tilde \varphi_\perp,
~\tilde \varphi_\parallel$ are twist $\tau=4$, and $\varphi_p(u)$ and 
$\varphi_\sigma(u)$ are the twist $\tau=3$ wave functions. 
Using Eqs. (8), (9) and (10), for the structure 
$\sigma_{\alpha\beta} \gamma_5$ we get 
\bea
\lefteqn{
T^{theor} =} \nnb \\
&& i \alpha \int dx \,e^{ipx}\Bigg\{ - 4 \frac{f_K}{2 \pi^2 x^4}
\ga \frac{\la \bar q q \ra}{12} + 
\frac{x^2 m_0^2}{192} \la \bar q q \ra \dr  
2 q_\beta x_\alpha \int du\,e^{-iuqx}\left[ \varphi_K(u) + x^2 \ga g_1(u) + 
G_2(u) \dr \right]\nnb \\
&&- 4 \frac{2f_K}{16 \pi^2 x^2} \ga \frac{\la \bar q q \ra}{12} +
\frac{x^2 m_0^2}{192} \la \bar q q \ra \dr
2 q_\beta x_\alpha \int du\, \int {\cal D} \alpha_i\, 
e^{-iqx(\alpha_1+u \alpha_3)} 
\left[\varphi_\parallel (1-2 u) - \tilde \varphi_\parallel \right]
\nnb \\
&&\mp \left[-\frac{4}{\pi^2 x^4} \ga \frac{\la \bar q q \ra}{12} + 
\frac{x^2 m_0^2}{192} \la \bar q q \ra \dr x_\alpha \right]
\left[ - q_\beta \int du \,e^{-iuqx} \left[ \varphi_K(u) + 
x^2 (g_1(u) + G_2(u) \right] \right] \Bigg\}~, \nnb \\ \nnb \\ \nnb \\
\eea
where 
\bea 
G_2(u) = - \int_0^u g_2(v) dv~. \nnb 
\eea
In deriving this equation we omit the terms which are equal to zero after
integration over $x$.
After Fourier transformation for the theoretical part of the correlator
function, we get
\bea
T^{theor} &=& - i \alpha f_K p_\alpha q_\beta \Bigg\{
4 \int du\, \varphi_K(u) \left[ \frac{\la \bar q q \ra}{12}
\ga \frac{2}{(p-qu)^2} + \frac{1}{2}\, \frac{m_0^2}{(p-qu)^4} \dr \right] \nnb \\
&+& \frac{8}{3} \la \bar q q \ra \int du\, \left[ (g_1(u) + G_2(u) \right]
\frac{1}{(p-qu)^4} \nnb \\
&-& \frac{2}{3} \la \bar q q \ra \int du\, \int {\cal D} \alpha_i
\left[\varphi_\parallel(\alpha_1,1-\alpha_1-\alpha_3,\alpha_3) (1- 2 u) -
\tilde \varphi_\parallel(\alpha_1,1-\alpha_1-\alpha_3,\alpha_3) \right]\nnb \\
&\pm& 4 \left[ \frac{\la \bar q q \ra}{12}  \int du\, \varphi_K(u)
\ga \frac{2}{(p-qu)^2} + \frac{1}{2}\, \frac{m_0^2}{(p-qu)^4} \dr \right] \nnb \\
&\pm& \frac{8}{3}  \la \bar q q \ra \int du\,  \left[ (g_1(u) + G_2(u) \right]
\frac{1}{(p-qu)^4} \Bigg\}~.
\eea
According to the general strategy of QCD sum rules, the quantitative
prediction for $g_{KYN}$ can be obtained by matching the representations of a
correlator (1) in terms of hadronic (Eq. (7)) and quark--gluon degrees of
freedom (Eq. (12)). Equating Eq. (7) and Eq. (12), and performing double
Borel transformation for the variables $p^2$ and $p_1^2$ in order to
suppress higher state and continuum contributions, we finally get the
following sum rules for $g_{K\Lambda N}$ and $g_{K\Sigma N}$ coupling
constants:
\bea
g_{K\Lambda N} \lambda_\Lambda \lambda_N &=& f_K M^2 
e^{(m_N^2+m_\Lambda^2)/2M^2} \sqrt{\frac{2}{3}} \nnb \\
&\times& \la \bar q q \ra \,
\Bigg\{ \frac{4}{3} \varphi_K(u_0) f_0 (s_0/M^2) +
\frac{1}{3 M^2}  \varphi_K(u_0) m_0^2  +
\frac{16}{3 M^2} \left[ g_1(u_0) + G_2(u_0) \right] \nnb \\
&+& \frac{2}{3 M^2} \int_0^{u_0} d \alpha_1 
\int_{u_0-\alpha_1}^{1-\alpha_1} \frac{d \alpha_3}{\alpha_3}\, 
\Bigg[ \ga 1 -2 \frac{u_0-\alpha_1}{\alpha_3} \dr
\varphi_\parallel(\alpha_1,1-\alpha_1-\alpha_3,\alpha_3) \nnb \\
&-& \tilde \varphi_\parallel(\alpha_1,1-\alpha_1-\alpha_3,\alpha_3) 
\Bigg] \Bigg\}~, \\ \nnb \\ \nnb \\  
g_{K\Sigma N} \lambda_\Sigma \lambda_N &=& f_K   
e^{(m_N^2+m_\Sigma^2)/2M^2} \sqrt{2} \nnb \\
&\times& \frac{2}{3} \la \bar q q \ra  \int_0^{u_0} d \alpha_1
\int_{u_0-\alpha_1}^{1-\alpha_1} \frac{d \alpha_3}{\alpha_3} 
\Bigg[ \ga 1 -2 \frac{u_0-\alpha_1}{\alpha_3} \dr
\varphi_\parallel(\alpha_1,1-\alpha_1-\alpha_3,\alpha_3) \nnb \\
&-& \tilde \varphi_\parallel(\alpha_1,1-\alpha_1-\alpha_3,\alpha_3) \Bigg]~,
\eea  
where the function 
\bea
f_n(s_0/M^2)=1-e^{-s_0/M^2}\sum_{k=0}^n \frac{(s_0/M^2)^k}{k!}~, \nnb
\eea
is the factor used to
subtract the continuum, which is modeled by the dispersion integral in the
region $s_1,~s_2 \ge s_0$, $s_0$ being the continuum threshold (of course
the continuum threshold for Eq. (13) is different than that for Eq. (14)),
\bea
u_0 = \frac{M_2^2}{M_1^2 + M_2^2}~,
~~~~M^2 = \frac{M_1^2 M_2^2}{M_1^2 + M_2^2}~, \nnb
\eea
and $M_1^2$ and $M_2^2$ are the Borel parameters. Since masses of $N$,
$\Lambda$ and $\Sigma$ are very close to each other, we can choose 
$M_1^2$ and $M_2^2$ to be equal to each other, i.e., $M_1^2 = M_2^2 =2 M^2$, 
from which it follows that $u_0=1/2$.  

From Eqs. (13) and (14) we see that the coupling constants $g_{K\Lambda N}$
and $g_{K\Sigma N}$ are determined by the quark condensate and wave
functions (for the structure $\sigma_{\alpha\beta} \gamma_5)$. 
We can deduce from these expressions that, the coupling
constant $g_{K\Lambda N}$ is determined mainly by the lowest twist 
$(\tau=2)$ wave function $\varphi_K(u)$, and the $g_{K\Sigma N}$ is
determined by the twist$(\tau=4)$ wave function, hence we expect that
$g_{K\Lambda N}>g_{K\Sigma N}$. Indeed our numerical calculations confirm this
expectation, as is presented in the next section.  

\section{Numerical analysis}

The principal nonperturbative inputs in the sum rules (13) and
(14) are the kaon wave functions on the light cone. In \cite{R3} a
theoretical framework has been developed to study these functions. In
particular, it has been shown that the wave functions can be expanded in
terms of the matrix elements of conformal operators which in a leading
logarithmic approximation do not mix under renormalization. For details, we
refer the reader to the original literature \cite{R4,R7}. In our numerical
analysis we use the set of wave functions proposed in \cite{R7}. The
explicit expressions of wave functions and the values of the various
parameters are:

\bea
\varphi_K(u,\mu) &=& 6 u \bar u \left[ 1 + a_2(\mu) C_2^{3/2} (2 u -1 ) +
a_4(\mu) C_4^{3/2} (2 u -1 ) \right]~, \nnb \\
g_1(u,\mu) &=& \frac{5}{2} \delta^2(\mu)\bar u^2 u^2 +
\frac{1}{2} \varepsilon(\mu) \delta^2(\mu) \Bigg[ u \bar u 
(2 + 13 u \bar u) \nnb \\
&& + 10 u^3 \ln u \ga 2 - 3u + \frac{6}{5} u^2 \dr +
10 \bar u^3 \ln \bar u \ga 2 - 3\bar u + 
\frac{6}{5}\bar u^2 \dr\Bigg]~, \nnb \\
G_2(u,\mu) &=& \frac{5}{3} \delta^2(\mu) \bar u^2 u^2~, \nnb \\
\varphi_\parallel(\alpha_i) &=& 120 \delta^2(\mu) \varepsilon(\mu)
(\alpha_1-\alpha_2) \alpha_1 \alpha_2 \alpha_3~, \nnb \\
\tilde \varphi_\parallel &=& - 120 \delta^2(\mu)
\alpha_1 \alpha_2 \alpha_3
\left[ \frac{1}{3} + \varepsilon(\mu) (1-3 \alpha_3) \right]~,
\eea
where the $C_2^{3/2}$ and $C_4^{3/2}$ are the Gegenbauer polynomials defined
as
\bea
C_2^{3/2} (2 u -1 ) &=& \frac{3}{2} \left[ 5(2 u-1)^2 +1 \right]~,\nnb \\
C_4^{3/2} (2 u -1 ) &=& \frac{15}{8} \left[ 21 (2 u -1 )^4 - 
14 (2 u -1 )^2 +1 \right] ~,
\eea
and $a_2(\mu=0.5~GeV)=2/3$ and $a_4(\mu=0.5~GeV)=0.43$. The parameter 
$\delta(\mu)^2$ was estimated from QCD sum rules  to have the value 
$\delta^2(\mu)=0.2~GeV^2$ \cite{R19}, $\varepsilon(\mu=1~GeV)=0.5$ \cite{R7}.
Furthermore we take $f_K = 0.156~GeV,~\mu_K (\mu=1 ~GeV) =
1~GeV,~m_0^2=0.8~GeV^2,~\la \bar q q \ra \ve_{\mu=1~GeV} = 
-(0.243~GeV)^3,~s_0^\Lambda = (m_\Lambda + 0.5)^2~GeV^2,~s_0^\Sigma = 
(m_\Sigma + 0.5)^2~GeV^2$. Also remember that all further calculations 
are performed at $u=u_0=1/2$.  

Having fixed the input parameter, one
must find the range of values of $M^2$ for which the sum rules (13) and
(14) are reliable. The lowest possible value of $M^2$ is determined by
the requirement that the terms proportional to the highest inverse power of
the Borel parameters stay reasonable small. The upper bound of $M^2$is
determined by demanding that the continuum contribution is not too large.
The interval of $M^2$ which satisfies both conditions is 
$1 ~GeV^2 < M^2 < 2~GeV^2$. The dependence of Eqs. (13) and (14) on
$M^2$ is depicted in Figs. 1 and 2. From these figures one can directly
predict
\bea
g_{K\Lambda N} \lambda_\Lambda \lambda_N &=& - 0.008  \pm 0.001~, \\
g_{K\Sigma  N} \lambda_\Sigma  \lambda_N &=& - 0.0006 \pm 0.0001~.
\eea

In determining the values of the strong coupling constants $g_{K\Lambda N}$
and $g_{K\Sigma N}$ we need the residues of hadronic currents,
i.e., $\lambda_N,~\lambda_\Lambda$ and $\lambda_\Sigma$, whose values are
obtained from the corresponding mass sum rules for the nucleon, $\Lambda$
and $\Sigma$ hyperons \cite{R17,R18}, as follows
\bea
\vel \lambda_N \ver^2 e^{-m_N^2/M^2} 32 \pi^4 &=& M^6 f_2(s_0^N/M^2) 
+ \frac{4}{3} a^2 ~,\\
\vel \lambda_\Lambda \ver^2 e^{-m_\Lambda^2/M^2} 32 \pi^4 &=& M^6
f_2(s_0^\Lambda/M^2) + \frac{2}{3} a m_s (1 - 3 \gamma) M^2 
f_0(s_0^\Lambda/M^2)\nnb \\ 
&&+ b M^2 f_0(s_0^\Lambda/M^2) + \frac{4}{9} 
a^2 (3+4 \gamma)~,\\
\vel \lambda_\Sigma \ver^2 e^{-m_\Sigma^2/M^2} 32 \pi^4 &=& M^6
f_2(s_0^\Sigma/M^2) - 2 a m_s (1+\gamma)M^2 f_0(s_0^\Sigma/M^2) \nnb \\
&&+b M^2 f_0(s_0^\Sigma/M^2) + \frac{4}{3} a^2~,
\eea
where 
\bea
a &=& - 2 \pi^2 \la \bar q q \ra ~, \nnb \\
b &=& \frac{\alpha_s \la G^2 \ra }{\pi} \simeq 0.12~GeV^4~, \nnb \\
\gamma &=& \frac{\la \bar q q \ra}{\la \bar s s \ra} - 1 = - 0.2~, \nnb
\eea
and the functions $f_2(x),~f_0(x)$ describe subtraction of the continuum
contributions, whose explicit forms are presented just after Eq. (14).
Dividing both sides of Eq. (17) $\lambda_\Lambda \lambda_N$ and 
Eq. (18) by $\lambda_\Sigma \lambda_N$, whose numerical values are 
obtained from Eqs. (19), (20) and (21), respectively, for 
$g_{K\Lambda N}$ and $g_{K\Sigma N}$ coupling constants we get
\bea
\vel g_{K\Lambda N} \ver &=& 10 \pm 2\nnb \\
\vel g_{K\Sigma N} \ver  &=& 0.75 \pm 0.15
\eea

Let us compare our predictions of 
$g_{K\Lambda N}$ and $g_{K\Sigma N}$ coupling constants with that of
traditional sum rules results for the structure 
$\sigma_{\mu\nu} \gamma_5 p_\mu q_\nu$ \cite{R14}. The results for 
these quantities in framework of the traditional QCD sum rules method are
\bea
\vel g_{K\Lambda N} \ver &=& 2.37 \pm 0.09~, \nnb \\
\vel g_{K\Sigma N} \ver &=&  0.025 \pm 0.015~.
\eea

When Eqs. (23) and (24) are compared, it is observed that the light cone
predictions on $g_{K\Lambda N}$ and $g_{K\Sigma N}$ are approximately
4 and 30 times larger, respectively, compared to that of the
traditional QCD sum rules results.

As an additional remark, it should be noted that the values of the coupling
constants $g_{K\Lambda N}$ and $g_{K\Sigma N}$ obtained in this work
differ from that of the exact SU(3)
prediction. Using de Swart's convention \cite{R20}, SU(3) symmetry predicts
\bea
g_{K\Lambda N} &=& - \frac{1}{\sqrt{3}} (3 - 2 \alpha_{\cal D} ) 
g_{\pi N N} ~, \\
g_{K\Sigma N} &=&  ( 2 \alpha_{\cal D} -1 ) g_{\pi N N} ~. 
\eea
Taking $\alpha_{\cal D} = 0.64$ \cite{R21} from Eqs. (24) and (25) we have
\bea
\vel \frac{g_{K\Lambda N}}{g_{K\Sigma N}} \ver \simeq 3.55~,
\eea
while our result for this ratio is 
$\vel g_{K\Lambda N}/g_{K\Sigma N} \ver \simeq 12$.

Finally we would like to state that, a more detailed analysis for
determination of the above--mentioned coupling constants from different
structures ($\gamma_5,\not\!P \gamma_5,\not\!q \gamma_5$) is needed. Such
an analysis may help to understand the source of discrepancy between
predictions of different structures.

\newpage
\begin{figure}
\vspace{25.0cm}
    \includegraphics{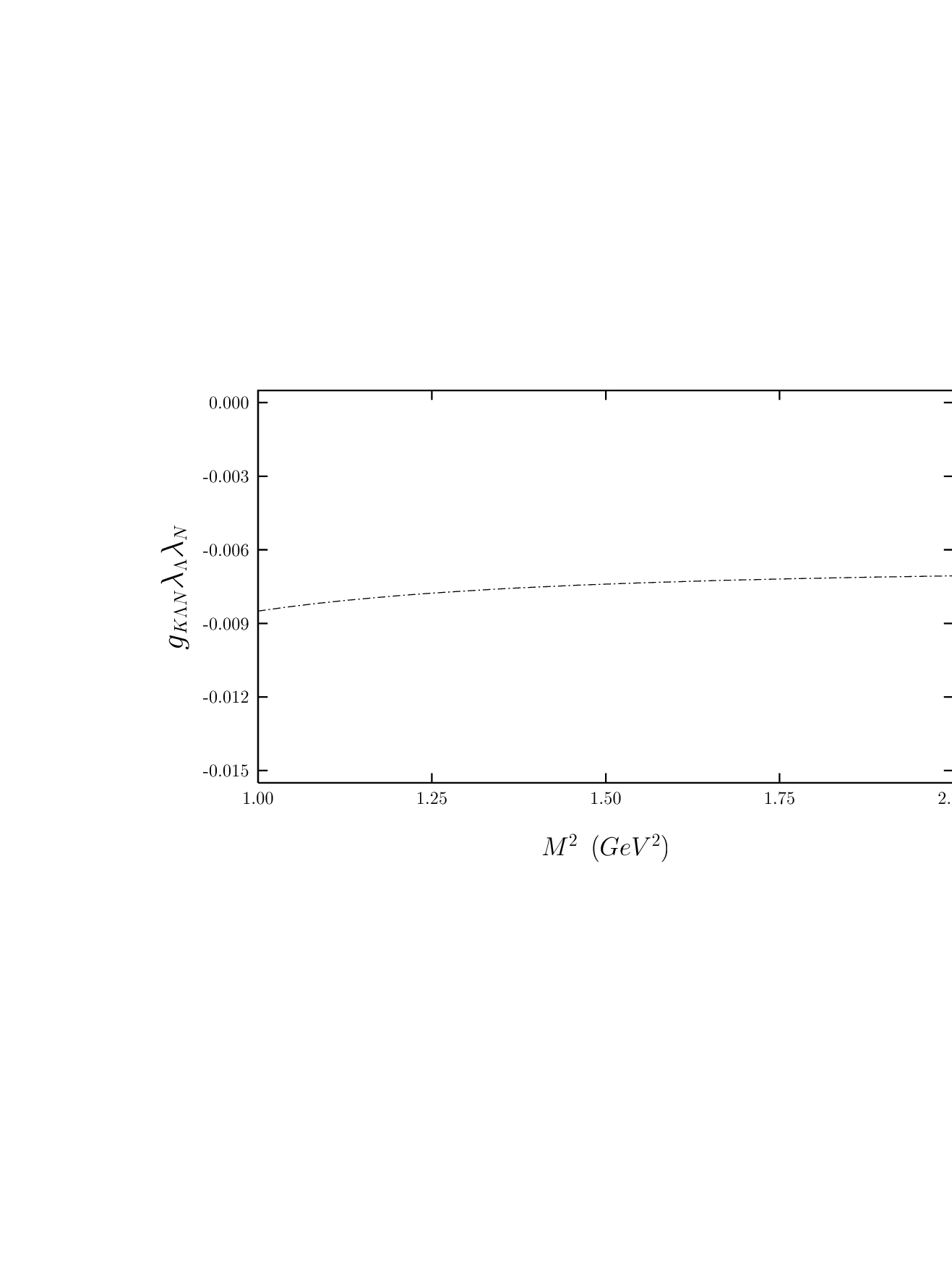}
\vspace{-17.cm}
\caption{ }
\vspace{16.5cm}
    \includegraphics{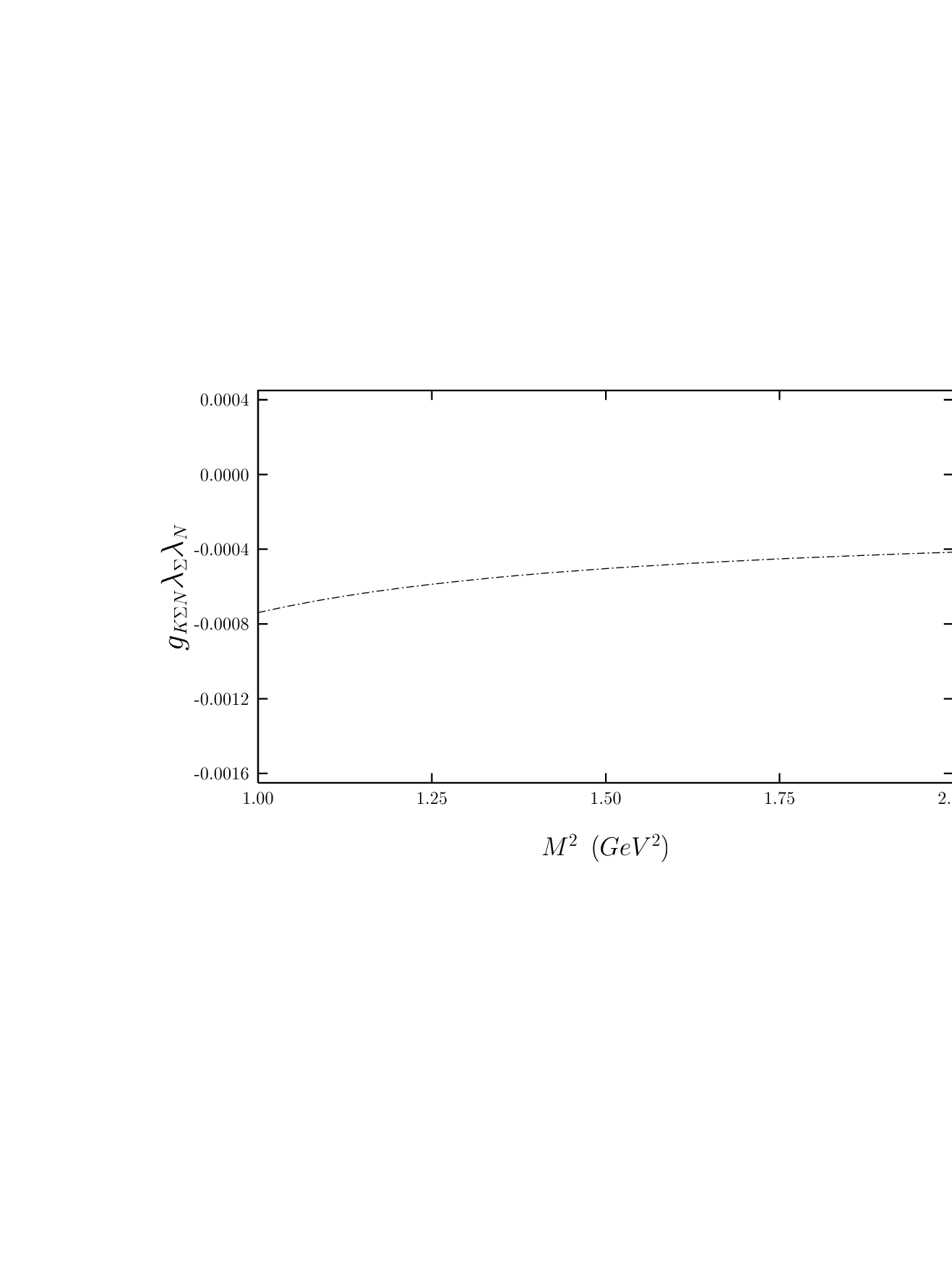}
    \vspace{.0cm}
\vspace{-5.cm}
\caption{ }

\end{figure}

\newpage
\section*{Figure captions}   
{\bf Fig. 1} The dependence of $g_{K\Lambda N} \lambda_\Lambda \lambda_N$
on the Borel parameter $M^2$. \\ \\
{\bf Fig. 2} The same as in Fig. 1, but for 
$g_{K\Sigma  N} \lambda_\Sigma  \lambda_N$.

\end{document}